\documentclass{PoS}
\usepackage{bbm}

\newcommand{\be}{\begin{equation}} 
\newcommand{\en}{\end{equation}}
\newcommand{\bea}{\begin{eqnarray}}
\newcommand{\ena}{\end{eqnarray}}

\newcommand{\hbo}{\hbox to 1 true cm {\hfill } }

\newcommand{\e}{\mathrm{e}}
\newcommand{\1}{\mathbbm{1}}

\title{String-like  theory as solution to the sign problem of a finite
  density gauge theory }

\ShortTitle{Z3 gauge theory with matter at finite densities}

\author{\speaker{Kurt Langfeld}  \\  
        Department of Mathematical Sciences, University of Liverpool,
        UK \\
        E-mail: \email{kurt.langfeld@liverpool.ac.uk}}

\abstract{
Z3 gauge theory with dynamical (bosonic) matter is studied in 4
dimensions with a finite chemical potential. This theory could be viewed
as an effective theory describing the centre vortex picture of QCD
colour confinement, but it is studied here with local interactions as
theory in its own right. It is shown that the sign-problem can be solved
by dualisation. The dual theory is derived: the pure gauge sector is a
theory of closed membranes with Nambu-Goto action, and matter is
described by open branes bounded by closed matter loops. The brane
theory is simulated with  Monte-Carlo techniques. Some evidence is
found that the theory possesses a weakly-renormalisable phase with
the scale set by a mass gap. Deconfinement at low temperatures and
finite chemical potentials appears as a percolation transition for
matter loops. 
}

\FullConference{XIII Quark Confinement and the Hadron Spectrum - Confinement2018\\
		31 July - 6 August 2018\\
		Maynooth University, Ireland}

\begin{document}

\section{Introduction}

The phasediagram of strongly interacting matter (as a
function of the temperature $T$ and the chemical potential $\mu $) is
believed to be informed by the transition from colour-confined to
deconfined matter. However, our current understanding of this
transition is derived from
effective considerations and model building that usually know little
about quark confinement. I give two examples:  to reconcile large
N-arguments with an 
effective quark model description led to the postulate of a
so-called quarkyonic phase~\cite{McLerran:2007qj} and, secondly,  effectice quark
models could justify the existence of ``chiral spirals'' or ``Fermi-Einstein
condensation''~\cite{Kojo:2009ha,Langfeld:2011rh}.

\medskip
By contrast, first principle lattice gauge simulations have shed light
on the QCD confinement mechanism for a vanishing chemical
potential. It started with the discovery of a linear-rising potential
between static quark sources induced by a colour-electric flux
tube forming between the sources~\cite{Bissey:2006bz}. It remained to
clarify why QCD electric flux tends to squeeze into tubes as opposed
to the case of flux spreading out in QED. A possible explanation was
offerred twenty years ago 
by Del Debbio, Faber, Greensite and Olejnik: after gauge fixing, the
projection of the SU(N) gauge field configurations to those taking
values in the centre $Z_N$ of the group produces a theory that retains
colour
confinement~\cite{DelDebbio:1996lih,DelDebbio:1998luz,Langfeld:2003ev}.
Further   significance was added by the discovery 
that the $Z_N$ gauge invariant degrees of freedom, the so-called
centre-vortices, have properties dictated by the physical mass scale
and survive in the continuum limit~\cite{Langfeld:1997jx}. A decade
long fruitful discussion followed which revealed the vortex signature
in the high temperature deconfinement
transition~\cite{Engelhardt:1999fd, Langfeld:2003zi}  or for chiral
symmetry breaking ~\cite{Gattnar:2004gx,Bowman:2010zr}, and
investigations of the vortex confinement mechanism extended to gauge
groups without a centre~\cite{Greensite:2006sm}. However, very little
is know about the centre vortex properties for light quark masses, 
let alone for finite density QCD.

\medskip  
Rather than adding to the extensive literature of effective descriptions of the
QCD phase diagram, I here would like to rise the question: Can we
deform QCD to a theory that still has linear colour confinement
and, at the same time, admits a first principle calculation of its
phase diagram?   

\medskip 
I will show that $Z_3$ gauge theory with bosonic matter in four
dimensions is an answer. At finite chemical potentials for the $Z_3$
matter, direct Monte-Carlo simulations are ruled out by a strong
sign-problem. The recent years have seen remarkable advances for
simulating those theories: Complexification of
fields~\cite{Sexty:2014dxa} gave rise to Complex Langevin 
simulations~\cite{Aarts:2012ft,Sexty:2016ard} or Lefschetz Thimble
inspired methods~\cite{Alexandru:2016ejd}. Algorithmic
advances~\cite{Langfeld:2016kty} may generically give reliable results
for medium size systems
(e.g.~\cite{Langfeld:2014nta,Garron:2016noc}). A powerful method for
solving sign problems is
dualisation~\cite{Endres:2006xu,Gattringer:2016kco}: a transformation
of field variables on the dual lattice may or may not produce a real
theory upon the integration of the original variables. This frequently
leads to non-local degrees of freedom sometimes called ``worms'' or 
flux-lines~\cite{Gattringer:2011gq}.

 \section{Brane theory from $Z_3$ gauge theory with $Z_3$
   matter}

 \subsection{The model}

 Let us consider a 4-dimensional hyper-cubic lattice of extent
 $N^3\times N_t$ and periodic boundary conditions. The protagonists of
 the lattice simulation are $Z_3$ group elements associated with the
 links of the 
 lattice (``Gluon'' fields) and with the sites (``matter'' field):
 \be
 U_\mu (x), \; \sigma (x) \; \in \; \{ 1,z,z^\dagger \} \; , \hbo
 z \; = \; \exp \left\{ i \, \frac{2 \pi}{3} \right\} \; .
 \label{eq:1}
\en
The partition function $Z$ of the theory features both, a pure gluonic
action and an interaction term:
\bea
Z &=& \sum _{\sigma, U_\mu } \; \exp \Bigl\{ S_g [U] \; +
\; S_f [\sigma,U] \, \Bigr\} \; ,
\label{eq:2} \\
S_g [U]  &=&  2 \beta \sum _{p}  \hbox{Re} \,  P_p \; , \hbo 
P_{p=(x,\mu < \nu)}  \; = \; U_\mu (x) \,
U_\nu (x+\mu) \, U^\dagger _\mu (x+\nu) \, U_\nu^\dagger (x) \;, 
\label{eq:3} \\
S_f [\sigma,U]  &=& 2 \kappa \sum _{x, \mu=1\ldots 3 } \, \hbox{Re} \, \Bigl[
\sigma ^\dagger (x) \, U_\mu (x) \, \sigma(x+\mu) \Bigr]
\label{eq:4} \\ 
&+& 2 \kappa \sum _{x, \mu=4 } \, \, \Bigl[ \; \e^\mu \, 
\sigma ^\dagger (x) \, U_\mu (x) \, \sigma(x+\mu)  \; + \; \e^{-\mu}
\, \sigma ^\dagger (x) \, U^\dagger_\mu (x-\mu) \, \sigma(x-\mu)  \;
\Bigr] \; , 
\nonumber
\ena 
where $p$ specifies an elementary plaquette of the lattice.

\subsection{Dualisation}

We are now going to sum over the fields $U_\mu , \sigma \in Z_3$. I
will only outline the calculation disregarding the matter fields by
setting $\kappa =0$. Further details will be presented
in a forthcoming publication. The $Z_3$ algebra greatly facilitates this calculation:
\be
U \in Z_3: \; \; \; U^3 = 1 \; , \; \; U U^\dagger =1 \; , \hbo \sum
_{U \in Z_3}  U =0 \; .
\label{eq:z}
\en
Because of these properties, we find for any bivariate function $f$,
which admits a Taylor expansion in both of its arguments:
$$
f(U,U^\dagger) \; = \; a \; + \; b \, U \; +\; c \, U^\dagger \; , 
$$
where $a,b,c$ are numerical constants.
In particular, the ``gluonic'' Gibbs factor can hence be written as:
\bea
\exp \{ S_g\}  &=& \prod _p \exp \; \Bigl[ \beta  (P_p + 
P^\dagger _p) \Bigr] \; = \; \prod _p c(\beta) \, \Bigl[ 1 + t(\beta )
\, (P_p+P_p^\dagger) \Bigr] \; .
\label{eq:5} \\
c(\beta) &=& \frac{1}{3} \, \e^{2\beta } \, + \, \frac{2}{3} \,
e^{-\beta } , \hbo
t(\beta ) \; = \;  \frac{ \e^{2 \beta } \, - \,  \e^{-\beta} }{ 
  \e^{2 \beta } \, + \, 2 \, \e^{-\beta} } \; .
\label{eq:6} 
\ena
Note that, for the interesting regime $\beta >0$, $t(\beta)$ is
positive. 

\medskip
The next step is to expand the brackets in (\ref{eq:5}) of the
product. For this purpose, we introduce auxiliary variables $n_p$ (one
for each plaquette $p$), which
later become the degrees of freedom that span the membrane:
$$
\1 \; + \; t \, P _p \; + \; t \, P_p^\dagger \; = \; \sum
_{n_p=-1,0,1} \Bigl[ \delta _{n_p,0} \, \1 \;+ \; \delta _{n_p,1} \, t\, P_p \;+
\; \delta _{n_p,-1} \, t \, P^\dagger _p \; \Bigr] \; = \; \sum
_{n_p=-1,0,1} \; t^{\vert n_p \vert }  \, P  ^{\, n_p} \; . 
$$
The gluonic Gibbs factor  (\ref{eq:5}) is then given by (with $V = N^3 \times
N_t$ the lattice size):
\be
\exp\{ S_g \} \; = \; c(\beta)^{6V} \; \sum _{\{n_p\}} \; \prod _p
\; t(\beta )^{\vert n_p \vert }  \, P  ^{\, n_p} \; .
\label{eq:7}
\en 
We have now a plaquette field $n_p$ at our finger tips: a
$n_p$-configuration has a value $0$,$1$ or $-1$ for each plaquette. If
we find $n_p=0$ for a given plaquette $p$, we say it is 
``trivial'', i.e., it does not contribute to the factor in
(\ref{eq:7}). We are now in the position to sum over all link fields
and to rewrite the partition function in terms of the new plaquette
field (recall $\kappa=0$):
$$
Z_g =  c(\beta)^{6V} \; \sum _{\{n_p\}} \; \sum _{\{U_\mu\}} \prod _p
\; t(\beta )^{\vert n_p \vert }  \, P  ^{\, n_p} \; .
$$
\begin{figure}[t]
  \includegraphics[width=15cm]{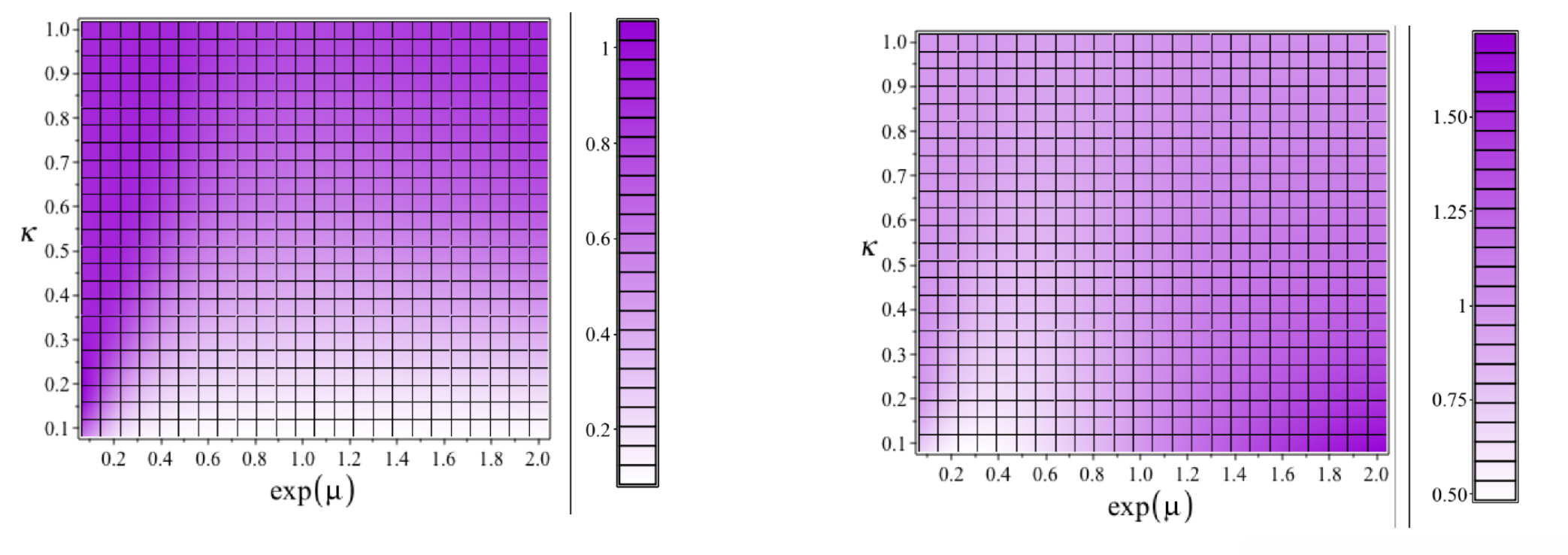}
  \caption{\label{fig:1}The emerging functions $t_f$ and $\Omega $ for
    a selected range of the hopping parameter $\kappa $ and the
    fugacity $\exp \mu$.}
\end{figure}
Depending on the plaquette configuration, $P  ^{\, n_p} $ is a product
of many link fields. Due to the property (\ref{eq:z}), many of this
products vanish upon summation over the $Z_3$ elements, e.g., if a
particular link $U_\ell$ of a non-trivial plaquette $p$ stands
alone. Another example is if two neighbouring non-trivial plaquettes
contribute each a factor $U_\ell $, we would find: $\sum U_\ell ^2
=0$. A way to avoid producing a vanishing contribution to the partition
function is the following: the non-trivial plaquettes only contribute
links $U_\ell $ and $U_\ell ^\dagger$ at the same time and we would
find $\sum U_\ell U_\ell ^\dagger = 3 $. We now see that the
summation over the $Z_3$ elements introduces constraints to set of
plaquette values $\{n_p\}$, and we introduce 
\be
\sum _{\{U_\mu\}} \prod _p \, P  ^{\, n_p} \; =: \; 3^{4V} \; \delta
_\mathrm{closed} (\{n_p\}) \; , 
\label{eq:8}
\en
where $\delta_\mathrm{closed} (\{n_p\}) =1$ if the contraints are
satisfied and vanishes in all other cases. I will now argue that if
the $n_p$ of a configuration $\{n_p\}$ form a set of closed oriented
surfaces, the constraint is satisfied. To this aim, let $c$ denote an
elementary cube of the lattice, and let $p \in c$ denote all
plaquettes forming the surface of the cube (the plaquette
contribute a $P_p$ pr $P_p^\dagger $ depending on the position on the
surface). The $Z_3$ Bianchi identity then yields:
$$
\prod _{p \in c } P_p [U_{\ell \in p}] \; = \; 1 \; . 
$$
Note that if a set of cubes has a plaquette $p$ in common, we can set
$n_p=0$ since the plaquettes come in pairs $P$ and $P^{\dagger
}$. Hence, if the set of $n_p$ form arbitrary but closed 
surfaces, we would find $\delta_\mathrm{closed} (\{n_p\})
=1$. Altogether, we find:
\be 
Z_g =  c(\beta)^{6V} \; 3^{4V} \sum _{\{n_p\}} \; \prod _p
\; t(\beta )^{\vert n_p \vert }  \, \delta_\mathrm{closed} (\{n_p\})
\; = \; c(\beta)^{6V} \; 3^{4V} \sum _{\{n_p\},\,  closed} \; \prod _p
\; t(\beta )^{\vert n_p \vert }
\label{eq:9}
\en
Defining the area $A$ of the closed surfaces and the ``surface
tension'' $\tau $ by,
$$
A[\{n_p\}] = \sum _p \vert n_p \vert \; , \hbo \tau := - \ln t(\beta )
  \; , 
$$
The gluonic partion function can be viewed as a theory of closed membranes
with a Nambu-Goto action:
\be
Z_g =  c(\beta)^{6V} \; 3^{4V} \sum _{\{n_p\}, \, closed } \; \exp \Bigl[
- \tau \, A[\{n_p\}] \Bigr] \; .
\label{eq:10}
\en

We now need to include the matter fields. The derivation follows the
lines above and starts with a character expansion of the Gibbs
factor. It starts noting that (for a given link $\ell = (x\mu)$) 
$K_\ell = \sigma ^\dagger (x) U_\mu (x) \sigma (x)$ is an element of
the group $Z_3$, and thus
\be 
\exp \{ S_f\} \; = \; \prod _\ell c_f (\kappa, \mu) \; \left\{ \1 \; +
  \; t_f (\kappa, \mu) \; \left[ \Omega (\kappa, \mu) \, K_\ell \; + \;
    \frac{1}{ \Omega (\kappa, \mu) } \, K^\dagger _\ell \right]
\right\} \; , 
\label{eq:20}
\en 
where a $c_f$, $t_f$ and $\Omega $ can be readily calculated. The
result is lengthy and will be presented elsewhere. I just point out
that for $\mu =0$, the action $S_f$ (\ref{eq:4}) is real, and we have
$\Omega (\kappa, 0)=1$. For the emerging string-like theory, it will be
important that $t_f$ and $\Omega $ are positive. A phenomenological
relevant range is $\kappa = 0.1 \ldots 1$ and $\exp \mu = 0.1 \ldots
2$. The colour plot in figure~\ref{fig:1} shows that both, $t_f$ and
$\Omega $, are indeed positive within safe margins. 

\medskip
\begin{figure}[t]
  \includegraphics[width=5cm]{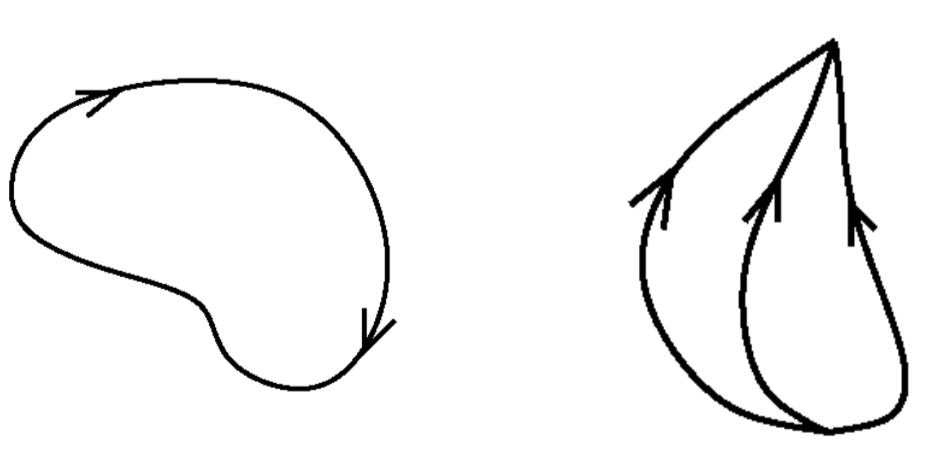} \hspace{1cm}
  \includegraphics[width=8cm]{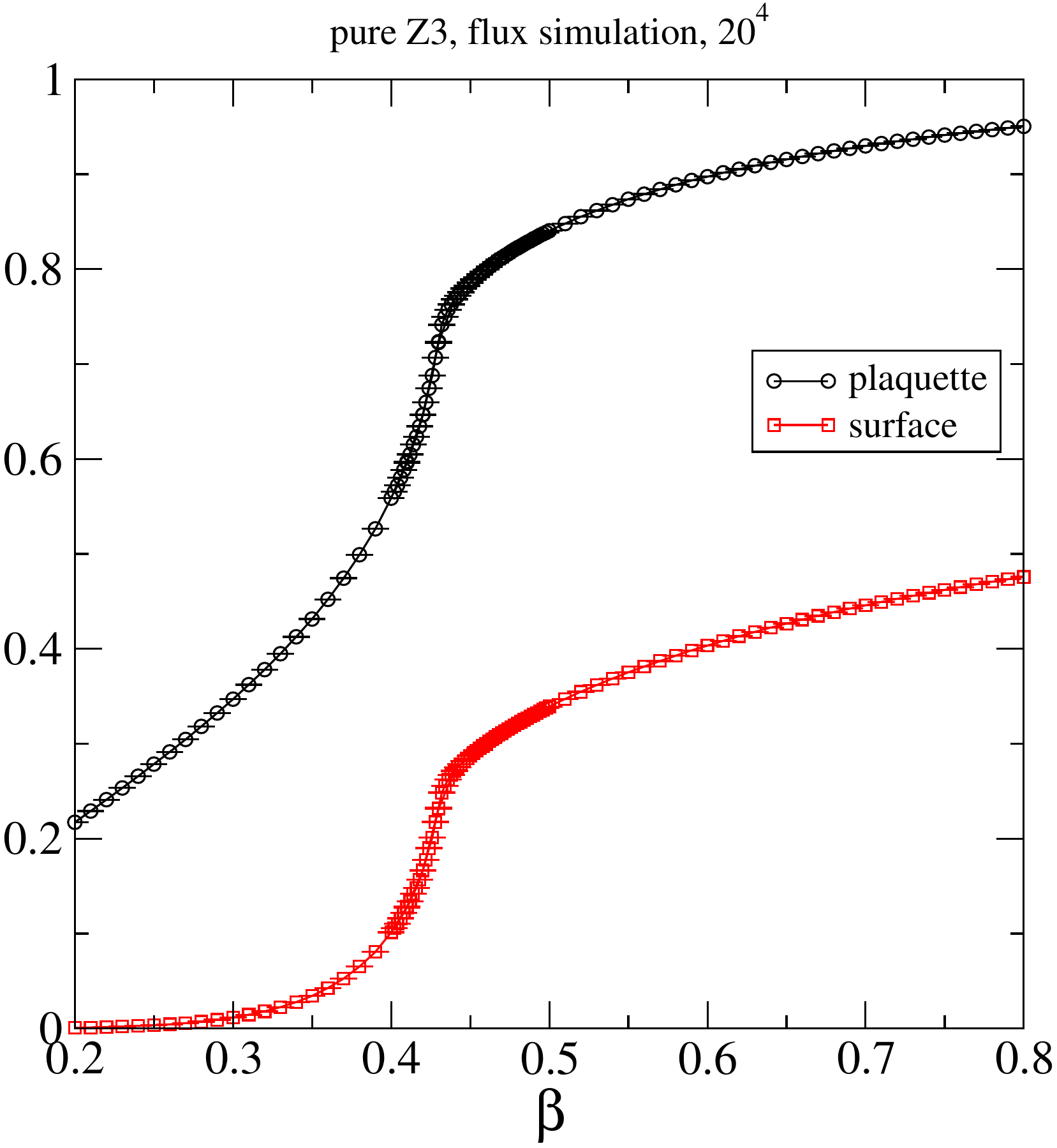}
  \caption{\label{fig:2} Left: Allowed flux lines configurations:
    ''mesonic'' type (left) and ``baryonic'' type (right).  Right:
    Average plaquette and the fraction of the 2-brane surface area as a
    function of $\beta $.}
\end{figure}
To expand the brackets in (\ref{eq:20}), we introduce for each link
$\ell $ a flux variable $k_\ell $ by
$$
\1 \; + \; \Omega (\kappa, \mu) \, K_\ell \; + \;
\frac{1}{ \Omega (\kappa, \mu) } \, K^\dagger _\ell  \; = \;
\sum _{k_\ell =-1,0,1} \Bigl[ \delta _{k_\ell,0} \, \1 \; + \;
\delta _{k_\ell,1} \, t_f \, \Omega \, K_\ell  \; + \;
 \delta _{k_\ell,-1} \, \frac{t_f}{\Omega } \, K_\ell \; \Bigr] \; . 
$$
This turns (\ref{eq:20}) into:
\be
\exp \{ S_f\} \; = \; c_f^{4V} \sum _{\{k_\ell\}} \prod _\ell\,
t_f^{\vert k_\ell \vert} \; \Bigl[ \Omega \, K_\ell \Bigr] ^{k_\ell}
\; . 
\label{eq:21}
\en 
The final step is to ``integrate out'' the gauge fields $U_\mu$ and
the matter fields $\sigma $. Inserting (\ref{eq:7}) and (\ref{eq:20})
into the partition function (\ref{eq:2}), the summation over the
original fields produces an equivalent formulation of the partition
function in terms of the plaquette variables $n_p$ and the flux
variables $k_\ell $:
\be
Z(\beta, \kappa, \mu ) \; = \; 3^{7V} c(\beta )^{6V} c_f(\kappa,
\mu)^{4V} \; \sum _{ \{n_p,k_\ell\}, \, closed }   t ^{A(n_p)} \; t_f
  ^{L(k_\ell)} \; \Omega ^{t_+ (k_\ell) - t_- (k\ell)} \; ,
  \label{eq:22}
\en 
where the constraints on the sets for $n_p$ and $k_\ell $ are as
follows:
\begin{itemize}
\item The set of oriented flux variables $k_\ell $ form closed lines
  that either form a loop or start (and end) in points of multiple of
  $3$ (see figure~\ref{fig:2} for an illustration). 
\item The set of oriented plaquette variables $n_p$ form either closed
  surfaces or open surfaces that are bounded by flux loops. 
\end{itemize}
Furthermore, $ A(n_p) $ is the total ``gluonic'' surface area, i.e.,
the number of non-trivial plaquettes $\sum _p \vert n_p\vert $. The
total length of the flux lines is denoted $L(k_\ell)$, and
$t_+(k_\ell) $ is the number of flux in positive time direction and $
t_- (k_\ell)$ the sum of those timelike fluxes that are oriented in
negative time direction. 

\section{ Brane theory - pure Z3 gauge theory }

\begin{figure}[t]
  \includegraphics[width=7cm]{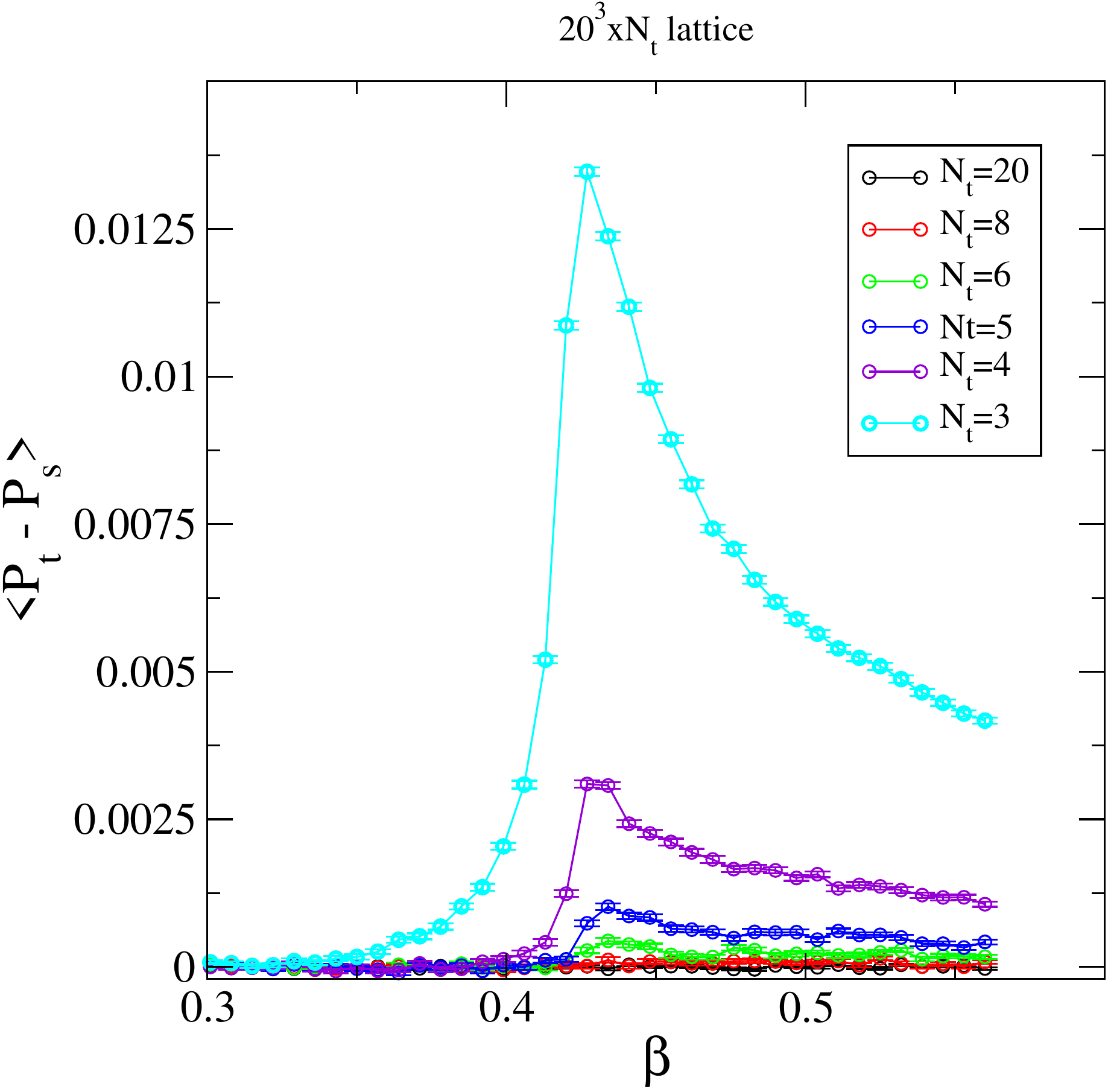} \hspace{1cm}
  \includegraphics[width=7cm]{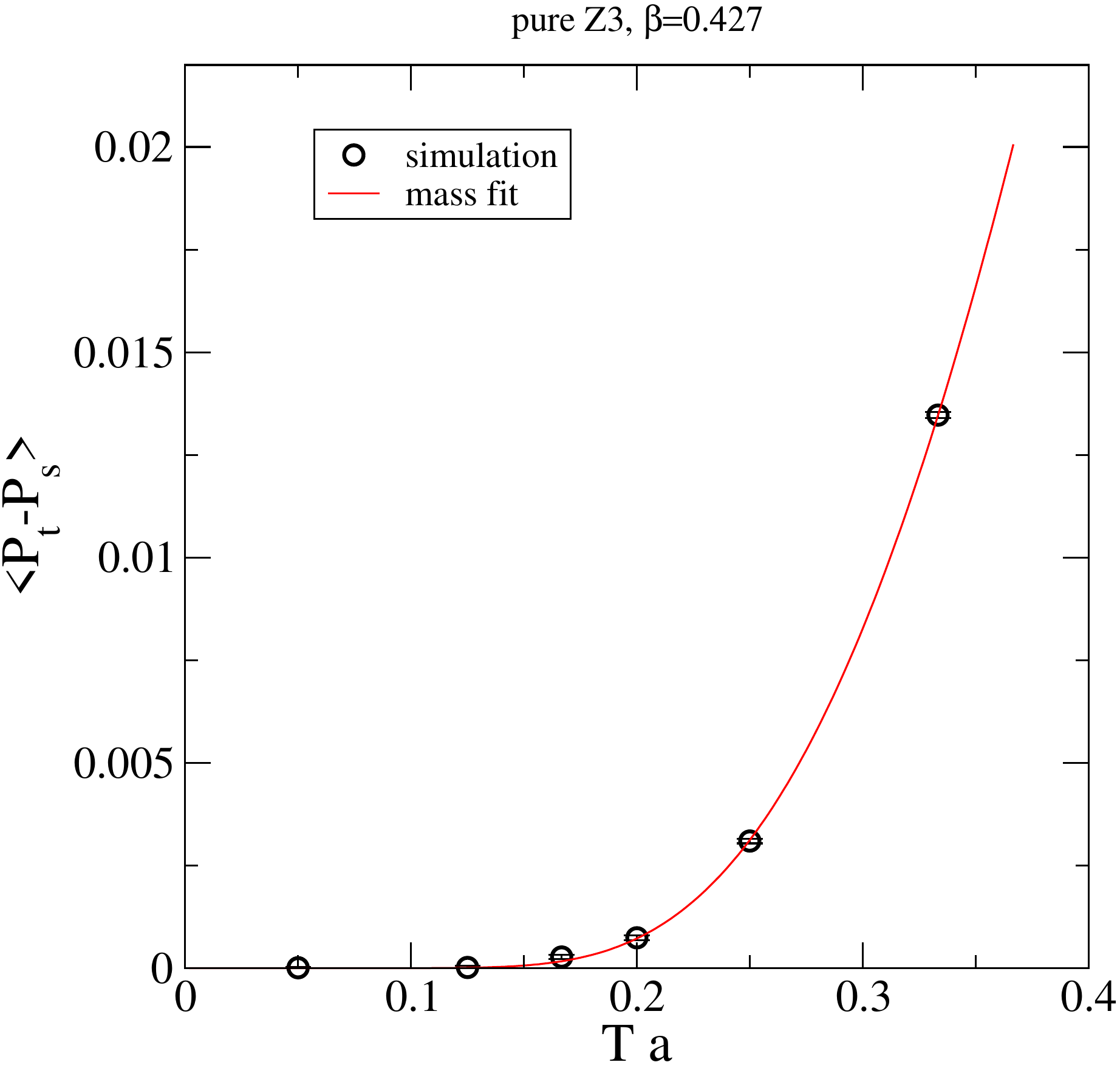}
  \caption{\label{fig:3} Left: Difference of time-like and space-like
    plaquettes on a $N^3 \times N_t$ asymmetric lattice as a
    function of $\beta $. Right: The difference for $\beta = 0.427 $
    as a function  of the temperature $T a = 1/N_t$. 
}
\end{figure}
In order to carry out a simulation of the brane theory
(\ref{eq:10}), we need to create configurations with closed
surfaces. I am using a standard Metropolis update that visits
elementary cubes on the lattice and deforms the existing configuration
of closed surfaces by adding and removing plaquettes of the surface
according to the orientation of the cube's plaquette. This
guarantees the MC update does not violate the constraints. The
underlying assumption is that this MC update is ergodic, i.e., that it
can generate {\it all} sets $n_p$ compatible with $\delta
_{closed}\not= 0$ in (\ref{eq:8}). At the current stage, the numerical
approach generating the specific sets of configuration should be
viewed as definition of the model until a rigorous link between the MC
approach and the Z3 theory is established. 

\medskip
As a first observable, I study the average plaquette
\be
\frac{1}{2} \frac{1}{6V} \; \frac{ \partial \ln Z_g }{\partial
  \beta } \; = \; \frac{1}{2} \,  \frac{ \partial \ln c(\beta)  }{\partial
  \beta } \; + \; \frac{1}{2} \, \frac{ \partial \ln t(\beta)  }{\partial
  \beta } \; \Bigl\langle \vert n_p \vert \Bigr\rangle \; , 
\label{eq:25}
\en
where I have used (\ref{eq:9}). 
The fraction $\langle \vert n_p \vert \rangle $ of non-trivial
surfaces on the lattice and the average plaquette are related by
numerical constants. Both quantities are shown in Figure~\ref{fig:2},
right panel. In the strong coupling regime at small $\beta $, the
2-brane surfaces are suppressed and the $\beta $-dependence of the
plaquette is dictated by the first term in (\ref{eq:25}). The
transition from the strong coupling regime to saturation at high
$\beta $ (sometimes called ``roughening transition'') is seen in the
brane theory as the onset when the brane surfaces start populating the
empty vacuum. 

\medskip
Temperature, say $T$, in quantum field theory is usually introduced by 
the extent of the torus in time direction:
$$
\frac{1}{T} \; = \; N_t \; a \, \hbo a: \; \hbox{lattice spacing}. 
$$
Renormalisation on the lattice is usually performed by demanding that
a physical mass $m$ does not change under a change of the lattice
regulator $a$. Measuring the scaling function $s(\beta) $ then defines
the lattice spacing as a function of the bare coupling parameter
$\beta $:
$$
m \, a \; = \; s(\beta) \hbo \Rightarrow \hbo a(\beta ) \; = \; \frac{
  s(\beta ) }{m} \; , 
$$
where $m$ plays the role of the free parameter (rather than the
coupling strength) and serves as a reference scale. A
correlation length (i.e., the inverse mass gap) is most easily detected
on asymmetric lattices using an observables that detects lattice
asymmetries. A possibility is the difference between time-like and
space-like plaquette averages, i.e., $\langle P_t - P_s \rangle $,
which  informs the entropy density in Yang-Millls theories. As soon
as the correlation length is large enough to equal the (smaller) time-like
extent $N_t a$ of the lattice, $\langle P_t - P_s \rangle $ will be
significantly different from zero. Figure~\ref{fig:3} shows the
difference as a function of $\beta $ for several sizes $N_t$. The data
show a clear peak around $\beta \approx 0.43$, which is more
pronounced for a small temporal extent. The right panel of the same
graph shows the difference for several temperatures at a fixed value
of $\beta $. If the theory possesses a mass gap $m$, thermal
excitations of the lightest excitation would imply:
$$
\langle P_t - P_s \rangle  \; \propto \; \exp \Bigl\{ - \frac{ m }{T}
\Bigr\} \; = \; \exp \Bigl\{ - \frac{ ma  }{Ta} \Bigr\} \; . 
$$
Indeed, the data are very well fitted by this ansatz (see red line if
figure~\ref{fig:3}). Repeating this analysis for several values for
$\beta $ determines the lattice spacing in terms of the mass gap $m$:

\medskip
\begin{tabular}{|l|llll|} \hline
  $\beta $ & 0.413 & 0.420 & 0.427 & 0.434 \cr \hline
  $m \, a(\beta )$ & 2.503(1) & 2.163(5) & 1.463(1) & 1.346(1) \cr \hline
\end{tabular}                                                                                      

\medskip
We see a UV type scaling: the lattice spacing $a$ shrinks with
increasing $\beta $. It is expected that the Z3 gauge theory possesses
a first order phase transition: the correlation length stays finite,
and a continuum limit $a \to 0$ does not exist. However, the model
could be {\it weakly renormalisable}: the infrared physics of the
theory is largely insensitive to the regulator for $a >
a_\mathrm{critical}$. A famous example of such a weakly renormalisable
theory is QED. Another quite recent example is from 
Philipsen and collaborators: they have started to study lattice QCD for
rather coarse lattices using the strong coupling expansion and manage
to extract rather robust infrared results (see
e.g.~\cite{Philipsen:2016wjt}).

\section{Branes, matter loops and finite densities}

\begin{figure}[t]
  \includegraphics[height=6cm]{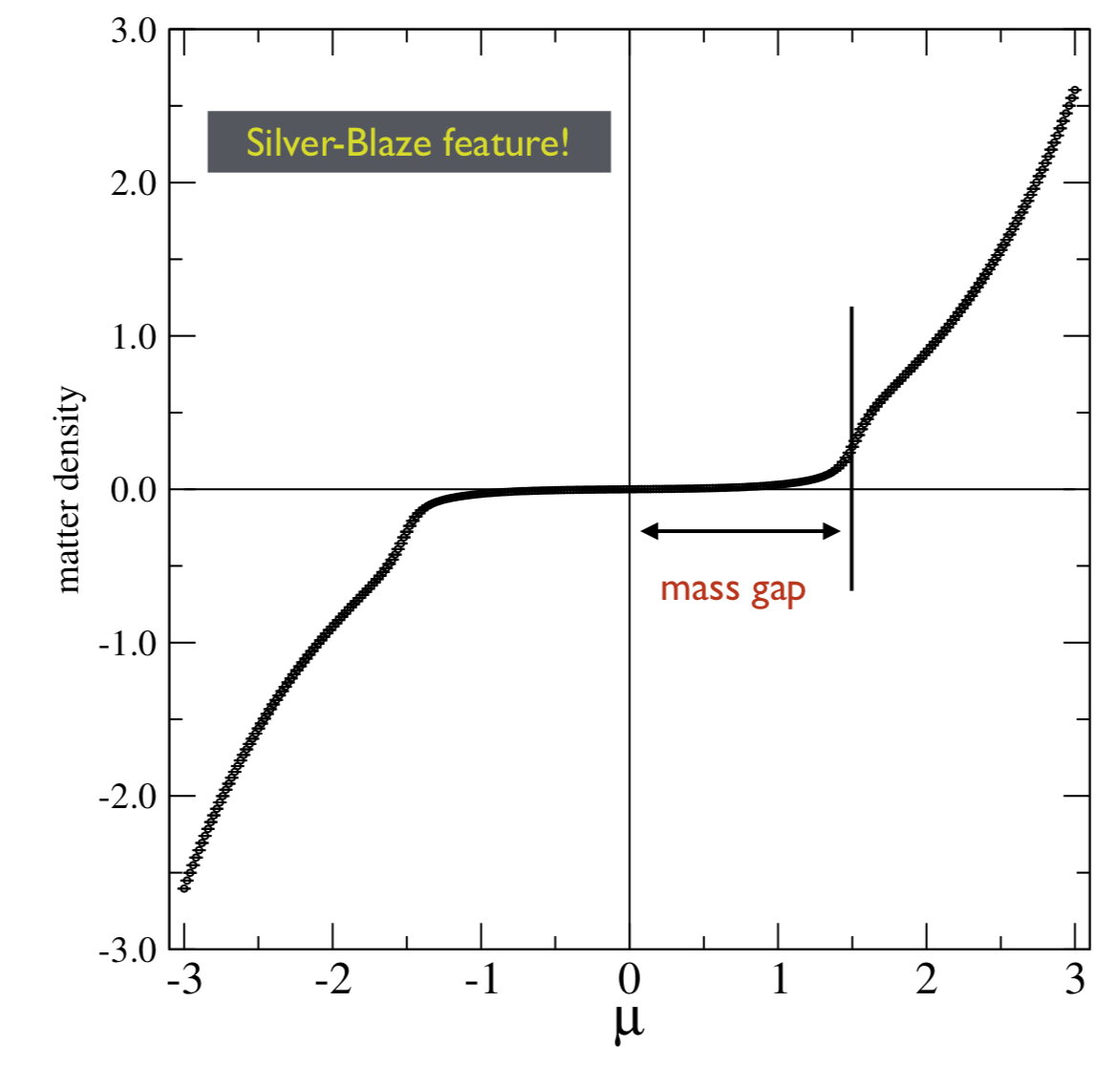} \hspace{1cm}
  \includegraphics[height=6cm]{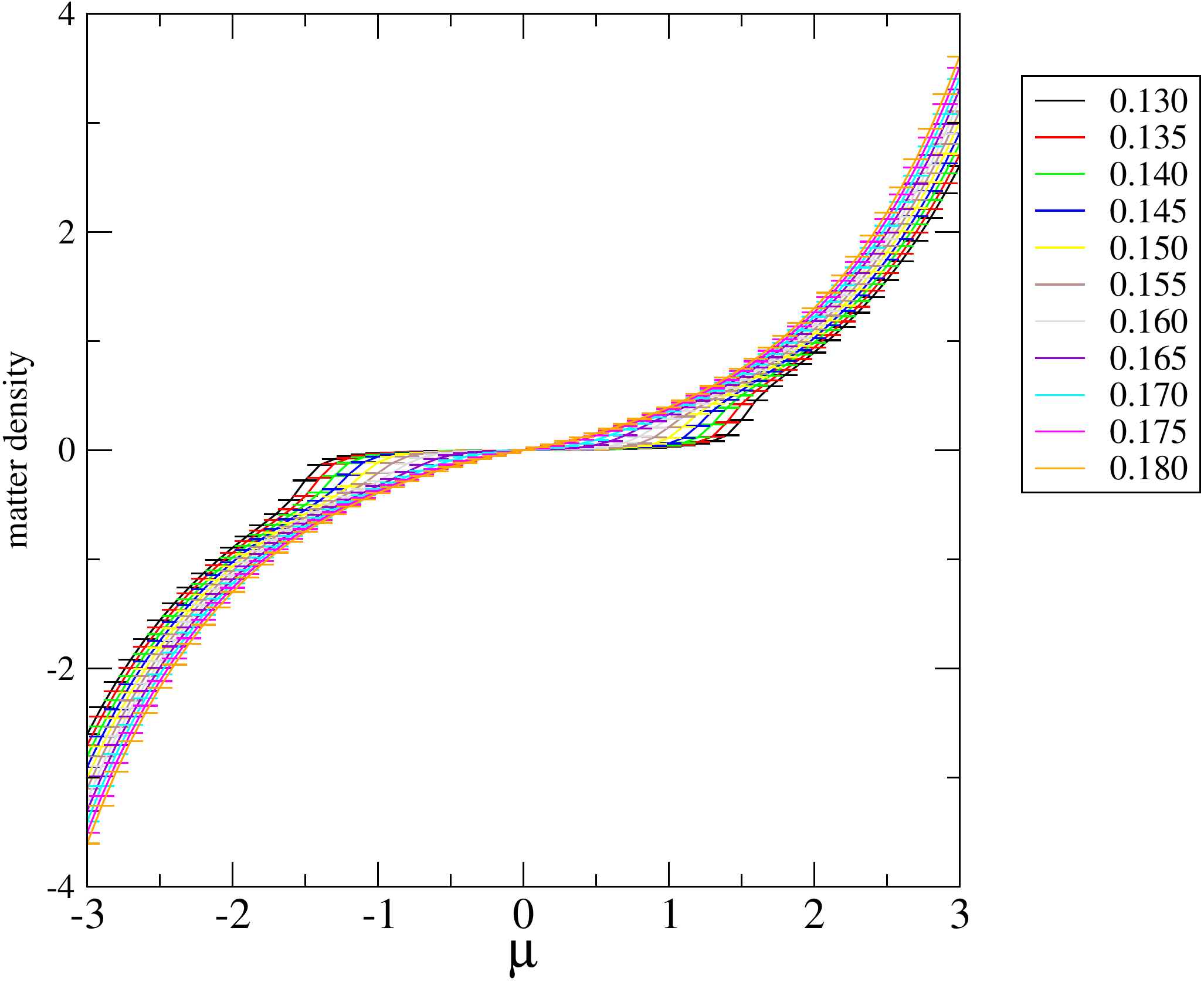} \hspace{1cm}
  \caption{\label{fig:4} Left: Matter density as a function of the
    chemical potential for $\beta = 0.43$ and $\kappa = 0.130$ and a
    $20^4$ lattice. Right: Same as left but for several values of
    $\kappa $. 
}
\end{figure}
We now consider the full Z3 gauge theory with dynamical matter at the
presence of a finite chemical potential $\mu $. As explained above,
the dual theory is a theory of closed and open branes bounded
by matter loops. The surface tension is informed by $t(\beta )$ and
the string tension of the matter loops by $t_f(\kappa, \mu)$ and
$\Omega (\kappa , \mu)$. The ``normalisation'' constants $c$ and $c_F$
also depend on the theory parameters:

\includegraphics[width=14cm]{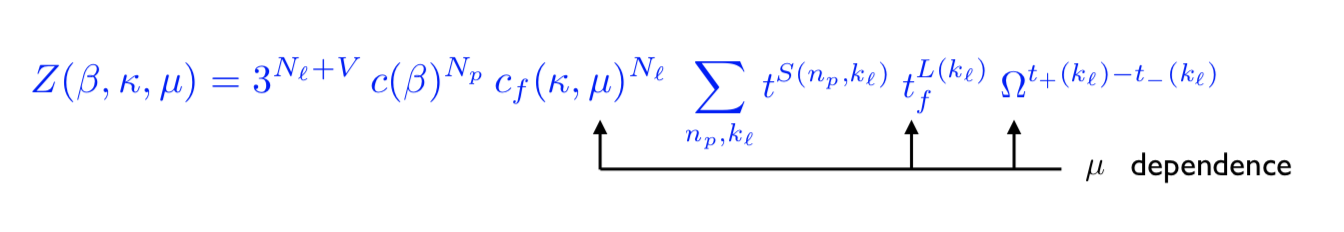}

\medskip
Note that the matter density
$$
\rho (\mu ) \; = \; \frac{ d \, \ln Z }{d \mu } 
$$ 
receives several contributions from the string theory variables:
\be 
\rho ( \mu ) \; = \; N_\ell \, \frac{ d \, \ln c_f  }{d \mu } \; + \;
\frac{ d \, \ln t_f  }{d \mu }  \Bigl\langle L(k_\ell)  \Bigr\rangle
\; + \; \frac{ d \, \ln \Omega }{d \mu }  \Bigl\langle t_+(k_\ell) -
t_-(k_\ell)  \Bigr\rangle \; .
\label{eq:30}
\en 
It is expected that for a chemical potential smaller than the matter
mass gap, $\mu < m_f$, the density remains zero in the infinite volume
limit (``Silverblaze'' feature). We stress that this is highly
non-trivial in the string theory formulation and involves intricate
cancellations in (\ref{eq:30}). At medium size chemical potentials,
the term on the far right of (\ref{eq:30}) strongly contributes to
the density. Note that closed matter loops, i.e., the ``mesonic'' 
loops, have $t_+(k_\ell) - t_-(k_\ell) $ and that only the ``baryonic''
loop configuration (see figure~\ref{fig:2}, left panel) do
contribute. 

\bigskip
\begin{figure}[t]
  \includegraphics[height=6cm]{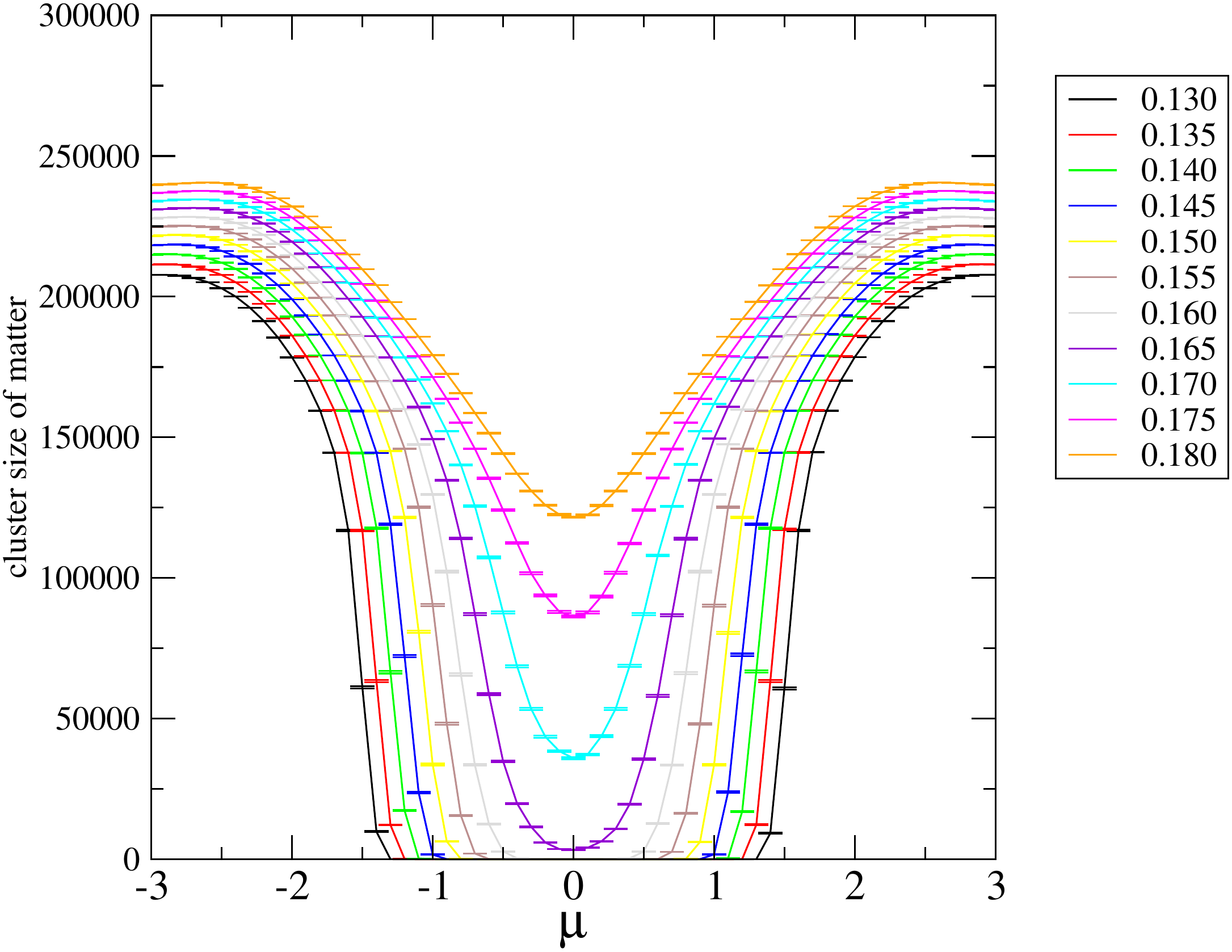} \hspace{1cm}
  \includegraphics[height=6cm]{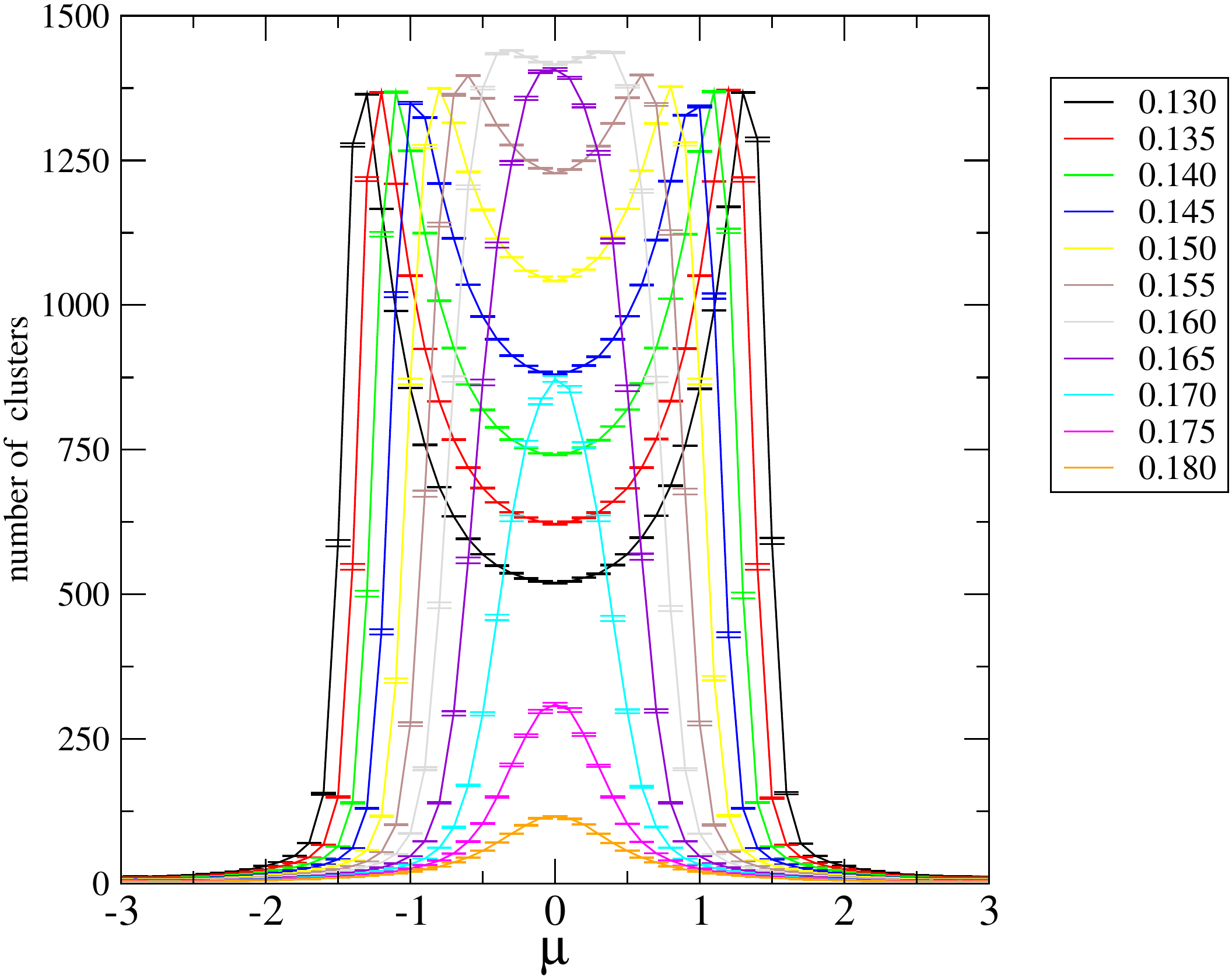} \hspace{1cm}
  \caption{\label{fig:5} Left: Average size of matter clusters as a
    function of the chemical potential. Right: average number of
    clusters populating the lattice. 
}
\end{figure}
We have simulated the full (brane) theory using a simple Metropolis
update. The elementary move for updating the closed ``gluon'' surfaces
is ``adding an elementary cube with oriented surfaces''. Two elementary
moves are necessary to update the open surfaces bound by matter
loops: A ``mesonic'' update adds the oriented loop around  an
elementary plaquette and the plaquette to the gluon surfaces; A
``baryonic'' update adding three matter lines part of a
$1\times 2$ rectangular surface. Changing the orientation of the
surface or the flux then generates the ``inverse'' move needed for
detailed balance. Details will be presented in a forthcoming
publication.

\bigskip  
Figure~\ref{fig:4} shows our numerical findings for the density $\rho
(\mu)$ as a function of the chemical potential $\mu $ for the case of
rather large matter mass dictated by a hopping parameter $\kappa =
0.130$. We indeed observe the ``silver blaze'' feature: the density
almost vanishes up to a certain threshold value $\mu _c$ when it
rapidly increases. The small variations of the density for $\mu < \mu
_c$ might be well explained by the finite lattice size. This
interpretation  is affirmed by reducing the matter mass, i.e.,
increasing $\kappa $ and observing that $\mu _c$ decreases (see
figure~\ref{fig:4}, right panel). For large enough $\kappa
\stackrel{>}{_\sim } 0.165 $, we
observe a immediate response of the density to the presence of a
chemical potential. We might interpret this value of $\kappa $ as the
massless case. 

\bigskip  
Let us explore further the properties of matter as a function of $\mu
$ and several masses informed by $\kappa $. Figure~\ref{fig:5}, left
panel, shows the average number of links that are connected to the
other links of one cluster. Let us focus on $\mu =0$: For large matter
masses (e.g., $\kappa =0.130$), we observe that only few and small
clusters are present. For $\kappa=0.165$, which we defined as the
massless limit above, we observe that the vacuum is populated by rather
large clusters.  Increasing $\kappa $ further, we
observe a proliferation of matter loops and, perhaps, a less
intersting phase for phenomenological purposes.

\medskip
If we now focus on the $\mu $-dependence, we observe for large matter
masses (e.g., $\kappa =0.130$) that, at the onset chemical potential, 
a proliferation of large clusters sets in. At the same time,  the
average number of clusters sharply drops (see figure~\ref{fig:5},
right panel). These findings are
compatible with the deconfinement mechanism put forward by Helmut Satz 
in the nineties~\cite{Satz:1998kg}: at a critical baryon density,
baryons start to overlap and the quarks can free percolate. 

\bigskip 
Note that there is also an important lesson to learn for QCD models:
in the subcritical region $\mu < \mu _c$ in the ``silverblaze 
region'', the properties of the (brane and loop) fields do
significantly depend on the chemical potential. These properties then
conspire and produce a cancellation of a $\mu $-dependence in physical
observables such as the density. This implies that effective models of
QCD (such as constituent quark models) need not necessarily be
independent from the chemical potential in the silverblaze regime.

\bigskip
{\bf Acknowledgements:} I thank David Schaich and Radu Tatar for
helpful comments, and, of course, the organisers for an excellent
conference. The numerical simulations were carried out at the HPC facility
Barkla at the University of Liverpool. Support is greatly
acknowledged.

\end{document}